\documentstyle[prd,aps,floats,epsfig]{revtex}  

\begin{document}  
\preprint{astro-ph/0006020} 
\draft

\renewcommand{\topfraction}{0.99} \renewcommand{\bottomfraction}{0.99}

\newcommand{\be}{\begin{equation}}
\newcommand{\ee}{\end{equation}}
\newcommand{\ba}{\begin{eqnarray}}
\newcommand{\ea}{\end{eqnarray}}
\newcommand{\rgl}{\rangle}
\newcommand{\lgl}{\langle}
\newcommand{\x}{\mbox{\boldmath $x$}}
\newcommand{\k}{\mbox{\boldmath $k$}}
\newcommand{\r}{\mbox{\boldmath $r$}}
\newcommand{\bu}{\mbox{\boldmath $u$}}
\newcommand{\nablab}{\mbox{\boldmath $\nabla$}}
\newcommand{\e}{\varepsilon}
\newcommand{\de}{\partial}
\newcommand{\nn}{\nonumber \\}
\newcommand{\lm}{{\ell m}}
\newcommand{\lmd}{{\ell' m'}}
\newcommand{\sqtpi}{\sqrt{\frac{2}{\pi}}}
\newcommand{\rhat}{\hat{\r}}
\def \pom {{\hspace{ -0.1em}I\hspace{-0.2em}P}}
\def \GeV {{\rm GeV}}
\def \MeV { {\rm MeV}}
\def \mb {{\rm mb}}
\def \mub {{\rm \mu b}}

\twocolumn[\hsize\textwidth\columnwidth\hsize\csname 
@twocolumnfalse\endcsname

\title{
%	{\it Submitted to Phys. Rev D}\\
%	\hspace{1.in}\\
	Perturbation Spectra in the Warm Inflationary Scenario}

\author{A.N. Taylor\footnote{email: ant@roe.ac.uk} \& 
	A. Berera\footnote{email: ab@ph.ed.ac.uk}}
 
\address{$^*$Institute for Astronomy, 
	Royal Observatory,
	Blackford Hill,
	Edinburgh, 
	EH9 3HJ, U.K. \\
	$^\dag$Department of Physics \& Astronomy, 
	University of Edinburgh, 
	Edinburgh, 
	EH9 3JZ, U.K. \\
	$^*$ant@roe.ac.uk, $^\dag$ab@ph.ed.ac.uk}

\maketitle

\begin{abstract}
We investigate the phenomenology of warm inflation and present
generic results about the evolution of the inflaton and radiation fields.
The general conditions required for warm inflation to take place are 
derived and discussed. 
A comprehensive approach is presented 
for the generation of thermally induced adiabatic and isocurvature 
perturbations and the amplitude of their spectra calculated. In addition 
we derive the ratio of tensor-to-scalar perturbations and 
find the spectral indices for adiabatic, isocurvature and tensor 
perturbations formed in the warm inflationary era.
These results are presented in a 
simplified and compact approach that is generally applicable.
 Our results are illustrated by inflation models with polynomial
and exponential potentials. We compare our analytic results
against numerical models and find excellent agreement. 
Finally, presently available data is used to 
put constraints on warm inflation and we discuss how near--future observations 
may distinguish the warm inflationary scenario from standard supercooled 
inflation. The main observable difference is the different scalar-to-tensor
ratio, and that the consistency relation between this and the slope of
tensor perturbations does not hold for warm inflation.

%\vspace{0.34cm}
%\noindent
PACS number(s): 98.80.Cq, 98.80.-k, 98.80.Es
%05.70.Ln, 11.10.Wx

\end{abstract}
\vskip2pc] 

%%%%%%%%%%%%%%%%%%%%%%%%%%%%%%%%%%%%%%%%%%%%%%%%%%%%%%%%%%%%%%%%
\section{Introduction}

One of the central paradigms in cosmology is the idea of inflation,
and its elegant solution to the problems of 
standard cosmology.
In addition inflation provides an explanation for the origin of structure
from quantum fluctuations frozen in by crossing the Hubble radius
during the rapid expansion. However in the standard inflationary 
scenario (see e.g. ref \cite{guth,ni,ci,olive,liddle}) the rapid 
expansion phase and subsequent 
reheating of the universe back to a radiation dominated model
are treated as two separate periods. The motivation for this 
was both simplicity and the belief that interactions with the 
field driving inflation would be minimal \cite{yoko}.
The switching on of field interaction dynamics at the end of 
inflation results in violent re- and pre-heating effects such as a resonant
amplification \cite{kof} which can lead to generic amplification
of perturbations \cite{bassett,brand,ep,llmw}. 

	From the point of view of particle physics, interactions are
inevitable, and they lead one to consider the effects on inflation of a
phenomenological ``friction'' term describing the decay of the 
inflaton field into other fields during inflation \cite{bf2,wi}.
One result of this is that density perturbations are no longer 
generated via quantum fluctuations, but are dominated by the larger thermal 
fluctuations \cite{bf2,abden}.

In this paper we consider the warm inflationary scenario \cite{wi,ab1}
from the 
phenomenological point of view.  Several types of
phenomenological warm inflation models exist
in the literature \cite{gmn,ramos,bellini,maia,ab2}.  
Warm inflation dynamics also has been studied in
quantum field theory \cite{bgr} and some of the
phenomenological models can be derived from first 
principles \cite{abden,bgr2,bk,bk2}. Our purpose here is to
determine generic observational features of the warm inflation
scenario, without referring to specific models.

The paper is organised as follows.
In Section II results are derive for evolution of the inflationary and
radiation fields during the slow-roll phase of warm inflation. 
Assuming the dissipation 
rate dominates over the expansion rate, a relationship is found between 
the radiation energy density and the energy density of the inflationary 
field. The spectrum of adiabatic perturbations is calculated in Section IIIA
and its spectral index in Section IIIB. Isocurvature perturbations are
also generated during warm inflation and we present a new
derivation of them in Section IIIC and of their spectral index in Section IIID.
Tensor perturbations are discussed in Section IIIE. We compare our 
results with standard inflation in Section IIIF. 
The conditions for ending warm inflation are discussed 
in Section IV.
Our results are 
illustrated by the specific cases of inflation with polynomial and power-law 
potentials in Section V.  Our analytic expressions are tested against
exact numerical models
in Section VI, where we find good agreement between the two.
In Section VII we discuss observational constraints
on warm inflation and how it may be distinguished from the standard inflation
scenario by future CMB experiments. Finally, our conclusions are presented
in Section VIII.  Two appendices are included in the paper. The first deals 
with a derivation of the thermal fluctuation of the inflationary 
field, while the second presents new results for  generic polynomial 
and exponential potentials.  Throughout this paper, the conventions
of \cite{liddlelyth} are followed.

\section{Evolution of the inflationary and radiation fields}

The evolution of the inflaton field, $\phi$, is given by
\be
	\ddot{\phi} + (3H+\Gamma)\dot{\phi} + V' =0,
\label{evolvphi}
\ee
where $H\equiv\dot{a}/a$ is the Hubble parameter, $a$ is the 
cosmological expansion factor, and $\Gamma$ is the 
dissipation coefficient. $V(\phi)$ is the potential of the inflationary field.
We assume a spatially flat universe.
 A dot denotes differentiation with time and a prime indicates 
$d/d\phi$. The energy--density of the radiation field, $\rho_\gamma$, 
obeys the transfer equation
\be
	\dot{\rho_\gamma} + 4 H \rho_\gamma =  \Gamma \dot{\phi}^2,
\label{evolvgam}
\ee
The equations are completed by the Friedmann equation for the 
evolution of the universe;
\be
	H^2 = \frac{8 \pi}{3m^2_p} (V + \rho_\gamma),
\label{fried_rad}
\ee
where $m_p$ is the Planck mass.
During an inflationary era the potential field dominates both the 
kinetic energy of the inflationary field and the energy density
of the radiation, so the Friedmann equation can be reduced to 
\be
	H^2 = \frac{8 \pi}{3m^2_p} V.
\label{friedmann}
\ee
We assume the slow--roll condition, $\dot{\phi}^2 \ll V(\phi)$, throughout 
the warm inflationary era, yielding
\be
	\dot{\phi} = - \frac{V'}{3H(1+r)},
\label{phidot1}
\ee
where the ratio of the production rate of radiation
to the expansion rate, $r$, is defined as  
\be
	r\equiv\frac{\Gamma}{3 H},
\label{rdef}
\ee
and parameterises the importance of dissipation in the warm inflationary 
model.

For the dissipation--dominated regime of warm inflation, the dissipation
rate, $\Gamma$, is much 
greater then the expansion rate; $r \gg 1$. The evolution of the 
inflaton field during this phase is governed by
\be
	\dot{\phi} = - \frac{V'}{3 r H}.
\label{phidot2}
\ee
Hence the slow-roll condition can be met even for relatively steep potentials, 
due to the damping effect of dissipation on the evolution of $\phi$.

If we also assume that during warm inflation radiation production is 
quasi-stable ($\dot{\rho_\gamma}\ll \Gamma \dot{\phi}^2$), equations 
(\ref{evolvgam}) and (\ref{phidot2}) give
\be
	\rho_\gamma \equiv  \alpha T^4
	=3 r \left(\frac{1}{2} \dot{\phi} \right)^2,
\label{radden}
\ee
where $\alpha\equiv g_*\pi^2/30$ is the Stefan-Boltzmann constant 
and $g_*$ is the number of degrees of freedom for the
radiation field. In the standard cosmology $g_* \approx 100$. 
Equation (\ref{radden}) shows that the energy density in the radiation field
is controlled by the kinetic energy of the inflationary field as well as
the dimensionless dissipation rate. As the kinetic energy of $\phi$ is
small during the slow-roll regime we require $r$ to be large to
generate substantial radiation energy-density. Equation (\ref{radden}) is not 
strictly true at the start and end of the warm inflationary phase,
as the stable radiation production condition will be violated.

In contrast the energy density of the inflaton during the slow-roll
phase is dominated by its potential energy, giving
\be
	\rho_\phi = V(\phi).
\label{inflden}
\ee
The radiation density can be re-written as
\be
	\rho_\gamma = \frac{1}{12 r} \left(\frac{V'}{H}\right)^2,
\ee
during warm inflation, since in the slow-roll phase the kinetic energy of the 
inflaton is controlled by the slope of the inflationary potential.

Introducing the dimensionless inflationary slow-roll parameter 
\be
	\varepsilon \equiv \frac{m_p^2}{16 \pi} \left( \frac{V'}{V}\right)^2
\label{defepsilon}
\ee
we can relate the energy density of the radiation field to the energy-density
of the inflaton by
\be
	\rho_\gamma = \left( \frac{\varepsilon}{2 r}\right) \rho_\phi.
\label{rad2phiden}
\ee
We have verified by numerical modelling (Section \ref{num}) that 
equation (\ref{rad2phiden}) holds very well during 
most of the warm inflationary phase.

We assume that warm inflation begins from a radiation dominated 
era, and starts when the energy density in the inflationary field
is equal to that in the radiation field. Initially the radiation 
energy is redshifted away by the quasi-exponential expansion but 
enters a quasi-static phase when radiation production stabilises
during the dissipation--dominated phase. 
Warm inflation ends when the universe heats up to become radiation dominated 
again, at a time when
\be
	\rho_\gamma = \rho_\phi.
\label{equal}
\ee
At this epoch the universe stops inflating and smoothly enters a 
radiation dominated big-bang phase. This is one of the most attractive features
of warm inflation, and suggests that its predictions are robust in the sense
that they are not altered during the smooth transition to the radiation-dominated
epoch. The material components of the
universe are created by the decay of either the remaining inflationary
field or the dominant radiation field.

The approximate relation, equation (\ref{rad2phiden}), and equation 
(\ref{equal}) suggest that an approximate condition for warm inflation 
to both begin and end is 
\be
	 \varepsilon =2 r,
\label{endwi}
\ee
although we do not expect this relation to be accurate
since the condition of stable radiation production is violated at both
epochs. In particular at the beginning of warm inflation the ambient radiation 
field is redshifted away during the sharp transition to warm inflation
and the radiation density drops rapidly. Then at the end of the warm inflation
phase, the production of radiation must end and we expect 
$\dot{\rho}_\gamma \approx 3 H \rho_\gamma \approx \Gamma \dot{\phi}^2$
during the turnover regime going from inflation-domination to 
radiation-domination.

In addition at both the beginning and end of warm inflation as the 
energy density of radiation is comparable to that of the inflationary 
potential, equation (\ref{fried_rad}) should be used to describe the 
evolution of the universe rather than equation (\ref{friedmann}),
introducing another approximation.

In Section VI we test the accuracy of our assumptions over the whole
course of a warm inflationary phase by numerically integrating 
equations (\ref{evolvphi}), (\ref{evolvgam}) and (\ref{fried_rad}) 
and find that for many models it appears 
that these approximations are accurate enough to be used to estimate 
the conditions over a large period of warm inflation, including the 
final few e-folds at the end of warm inflation.

The condition for a warm inflationary phase can be summarised by
\be
	\e < 2 r.
\label{wicond}
\ee
It is straightforward to show that the same result can be derived from 
the requirement of accelerated expansion in the warm inflationary epoch.
This requires that $\ddot{a}>0$ implying $-\dot{H}/H^2<1$ or
$-\dot{\phi}H'(\phi)/H^2<1$. Using equations 
(\ref{friedmann}) and  (\ref{phidot2})
we again find the inequality of equation (\ref{wicond}).
The slow-roll condition and the additional strong dissipation condition
\be
	r > 1
\ee 
then implies warm inflation. One immediate consequence is that the usual 
inflationary condition, $\e<1$, is no longer required as inflation 
(quasi-exponential expansion) can be maintained for steep potentials
due to the damping of dissipation.

Finally, one can differentiate equation (\ref{phidot2}) with respect to 
time and, following from the slow-roll condition $\ddot{\phi} \ll V'$,
we find a constraint on the second derivative of the potential,
\be
	|\eta | \ll 3 r^2
\ee
where 
\be
	\eta \equiv \frac{m_p^2}{8 \pi} \frac{V''}{V}
\ee
is the curvature slow-roll parameter for inflation. This should be 
regarded as an extra condition for sustained warm inflation, and is
a generalisation of the usual inflationary constraint, $|\eta | \ll 1$.
Hence warm inflation can take place for a wider range of potentials
than standard supercooled inflation.

\section{The perturbation spectra}

\subsection{Adiabatic perturbations}

Density perturbations are calculated in the same way as for standard inflation
\cite{bf2,abden},
via the equation 
\be
	\delta_H = \frac{2}{5}
	\left( \frac{H}{\dot{\phi}} \right) \delta \phi,
\label{adpert}
\ee
where $\dot{\phi}$ is given by equation (\ref{phidot2}). This equation 
holds for a mixture of inflationary  and radiation fields, in the 
quasi-static regime of radiation production. It also can be shown that 
this result holds for any scalar field interacting adiabatically with another field,
in the quasi-static limit, 
with the equation of state $p = \omega \rho$. In the presence of non-adiabatic
perturbations (see Section \ref{isocurv}) this relation does not necessarily hold. 
Specifically, if there non-adiabatic pressure perturbations, these can
induce curvature perturbations to evolve on super-horizon scales \cite{wands}.
However, we expect these effects to be small for the scenario discussed here
and equation (\ref{adpert}) can still be used for the order-of-magnitude estimates 
presented here.

The amplitude of perturbations, $\delta_H$, is related to the final 
dimensionless linear power spectrum of mass perturbations by
\be
	\Delta^2(k) = \left(\frac{k}{aH}\right)^4 \! T^2(k)\,\delta_H^2 (k)
\ee
where $T(k)$ is the matter transfer function relating initial spectra
to the final, post-recombination linear spectra.

The fluctuations in the inflationary field are no longer generated 
by quantum fluctuations, but by thermal interactions with the radiation 
field;
\be
	\delta \phi^2 = \frac{k_F T}{2 \pi^2},
\label{dphi}
\ee
where $k_F=\sqrt{\Gamma H}=\sqrt{3r}H\ge H$ is the freeze-out scale at which 
dissipation damps out the thermally excited fluctuations.
The origin of the freeze-out wavenumber $k_F$ is explained
in Appendix A based on the original derivation in \cite{abden}.
Note that in the 
warm inflationary epoch, the freeze-out length scale is generally
much smaller than the 
Hubble radius.  As one final point,
when $\Gamma\le H$ and $T\le H$, the amplitude of
thermal fluctuations is equal to the amplitude of quantum fluctuations 
generated by Hawking radiation.

Combining equations (\ref{friedmann}), (\ref{phidot2}), (\ref{adpert}) and 
(\ref{dphi}) we find
\be
	\delta_H^2 =  \frac{32}{75  m_p^4} 
	\frac{V}{\varepsilon} \left[r^{5/2}
	\left( \frac{\sqrt{12}T}{H}\right)\right] .
\ee
The temperature of the radiation field is related to the potential by
equations (\ref{radden}), (\ref{inflden}) and (\ref{rad2phiden}) yielding 
\be
	T^4=\frac{\varepsilon V}{2 r \alpha}.
\ee
Expressing the amplitude of density perturbations in terms of 
inflationary parameters we find
\be
	\delta_H^2 = \frac{16}{25\sqrt{ \pi} m_p^3}  
\left(\frac{2}{\alpha} \left( \frac{r^3 V}{\varepsilon}\right)^3\right)^{1/4}.
\ee
The ratio of thermal to quantum fluctuations is $r^{5/2}T/H$. During 
strong dissipation, $r \gg 1$, while towards the end of the warm inflationary 
era $T/H \approx m_p/V^{1/4}\gg 1$. Hence the thermal spectrum
will dominate over the quantum spectrum of fluctuations during a warm 
inflationary phase.

\subsection{Spectral index of adiabatic perturbations}

The spectral index of the adiabatic scalar perturbations is given by
\be
	n_s-1 = \frac{d\ln \delta^2_H}{d\ln k }
\ee
where the logarithmic interval in wavenumber is related to the 
number of e-folds from the start of inflation, $N(\phi)$, and the scalar 
field by
\be	
	d \ln k(\phi) = -d N(\phi) = -\frac{8 \pi}{m^2_p} \frac{rV}{V'} d \phi.
\ee
Hence the spectral index of adiabatic density perturbations in warm 
inflation is
given by
\be
	n_s = 1 - \frac{3}{4 r} (3 \varepsilon - 2 \eta).
\label{slopewi}
\ee

As we require that $r\gg 1$ we expect the deviation from 
scale invariance to be suppressed in the warm inflationary model, until
$\e\approx r$, at the end of the warm inflation phase.
If $\varepsilon > (2/3) \eta$, or $\eta$ is negative, the 
spectrum of adiabatic fluctuations will be 
red, with the spectrum suppressed on small scales, as in standard 
supercooled inflation which generally also
produces slopes $n_s \le 1$. 
However, in the opposite case,
when $\varepsilon < (2/3) \eta$, the 
spectrum of adiabatic fluctuations will be 
blue, in contrast with predictions from generic
standard inflation models.
With a blue spectrum, the interesting possibility exists of producing
primordial black holes
on small scales \cite{blue}.

\subsection{Isocurvature perturbations}
\label{isocurv}

Isocurvature perturbations also are generated in the warm inflationary era due
to thermal fluctuations in the radiation field. These perturbations
can be characterised by fluctuations in the entropy, $S_X$, where 
the species $X$ is undergoing thermal fluctuations relative to the 
number density of photons
\be
	S_X= \frac{3}{4} (\delta_X - \delta_\gamma)
\ee
and we have assumed the $X$-field is the ambient radiation field
excluding photons. 

Assuming that the isocurvature fluctuations are thermal in origin,
their perturbation spectrum is
\ba
	\delta^2_{\rm iso} &=& \frac{1}{25 m_p^2}\frac{k_F T}{2 \pi^2} \nn
%	\approx r^{1/2} \frac{H^2}{m^2_p} \left(\frac{T}{H}\right)
	&=& 
	\frac{1}{25 \sqrt{\pi^3}}
	\left( \frac{2 r \varepsilon}{\alpha}\right)^{1/4} 
	\left( \frac{V^{1/4}}{m_p}\right)^3.
\ea
This relation is true only for non-interacting fields, but again should 
hold for order-of-magnitude estimates.
The ratio of isocurvature to adiabatic perturbations is 
\be
	R_{\rm iso} \equiv 
	\frac{\delta^2_{\rm iso}}{\delta^2_H} = \frac{\varepsilon}{16 \pi r^2}
\ee
so that when $r\gg1$ and $r \ge \varepsilon$  adiabatic perturbations will 
dominate.
In practice the scalar modes will form a quadratic sum of adiabatic and
isocurvature modes,
\be
	\delta_s^2 = \delta^2_H + \delta^2_{\rm iso} ,
\ee
and which can be separated observationally since isocurvature generated
fluctuations of the CMB are out of phase with adiabatic perturbations.
Another discriminator is if super-horizon
perturbations can be observed, for instance in the CMB \cite{bfh}. Since 
isocurvature modes do not contribute to curvature perturbations 
(in the comoving gauge) the contribution from isocurvature
modes should be absent from super-horizon perturbations, leading to 
a drop in amplitude of the scalar perturbations at the horizon 
scale. Isocurvature perturbations have previously been discussed
in the context of warm inflation by \cite{fl}, although no detailed
calculation of their amplitude was given.

\subsection{Spectral index of isocurvature perturbations}
 Following the derivation of the spectral index of adiabatic perturbation,
the spectral index of the isocurvature perturbations, defined by
\be
	n_{\rm iso} \equiv \frac{d \ln \delta^2_{\rm iso}}{d \ln k },
\ee
 is 
\be
	n_{\rm iso} = -\frac{1}{4 r} (\varepsilon + 2 \eta).
\ee
This differs from isocurvature fluctuations generated 
by quantum fluctuations, where $n_{\rm iso} = -2 \varepsilon$. If 
$r$ is large, the spectrum will again tend to be flat, until $\e \approx r$.  
If $\varepsilon > -2\eta$
the thermal isocurvature spectrum will be red as it is in
standard supercooled inflation.

\subsection{Tensor perturbations}

Tensor perturbations do not couple strongly to the thermal background and
so gravitational waves are only generated by quantum fluctuations, as
in standard supercooled inflation. Their spectrum is given by 
\cite{liddlelyth}
\be
	A^2_{\rm g} = \frac{32 V}{75 m_p^4},
\ee
with spectral index
\be	
	n_{\small\rm g} \equiv \frac{d \ln A^2_g}{d \ln k} = -  \frac{2\e}{r},
\ee
and the inequality $-4<n_g<0$ arising from the warm inflation condition
$\e \le r$.

The ratio of tensor to scalar perturbations is 
\be
	R_g \equiv \frac{A^2_{\rm g}}{\delta_H^2} =
	 \frac{\sqrt{\pi}}{3} \left(\frac{2\varepsilon}{ r^3} \right)^{3/4}
	\left(\frac{\alpha V}{m_p^4} \right)^{1/4}.
\ee
Hence, as we are in the strongly dissipative regime, where $r\gg 1$ and
$\e \le r$, we expect the contribution from gravitational waves to be small
unless the magnitude of the potential is large or $r$ is order unity.

The ratio of the amplitude of isocurvature to tensor perturbations 
scales as 
\be
	\frac{\delta^2_{\rm iso}}{A_g^2} =
	\frac{3 m_p}{32\sqrt{\pi^3}}
	\left(\frac{2 r \e  }{\alpha V}\right)^{1/4},
\ee 
which 
will be greater than unity unless the scale of the potential is large.
Hence we can expect that the inequalities
\be
	R_g < R_{\rm iso} < 1
\label{ratioineq}
\ee
in general will hold. We find that this is true for all the of the 
numerical models we consider in Sections V and VI.

\subsection{Comparison with Standard Supercooled Inflation} 

A nice feature of warm inflation is that these results are significantly
different from standard supercooled inflation case, where
\ba
	 \delta_H^2 &=& \frac{32V}{75 m_p^4\varepsilon}, \nn
		n_s &=& 1  -6 \varepsilon +2 \eta, \nn
		n_g &=& -2 \e, \nn
		  R_g &=& \varepsilon.
\label{slopesi}
\ea
Thus near future observations should be able to distinguish between 
the two. An immediate consequence of our analysis is that there need
be no consistency relationship between the scalar spectral index and 
the ratio of tensor to scalar perturbations, as in standard inflation.
We discuss the observational differences between the models in Section VII.

\section{Conditions at the end of inflation}
At the end of the warm inflationary epoch, the dissipation parameter, $r$,
is related to the slow-roll parameter, $\e$, by equation (\ref{endwi}).
Substituting in equations (\ref{rdef}), (\ref{friedmann}) and (\ref{defepsilon})
we form a differential equation which can be solved for $V(\phi)$ to find the 
magnitude of the potential field at the end of warm inflation,
\be
	V(\phi_{\rm end}) = \frac{\pi}{6} 
	\left(\frac{\Gamma}{m_p}\right)^2 \phi_{\rm end}^4,
\ee
up to a constant, but
independent of the shape of the inflationary potential.
This selects out the $V \approx \phi^4$ potential as a special case
where warm inflation never ends (see Appendix B). This is due to 
the energy density of the radiation field ``tracking'' the energy 
density in the inflationary field, $\rho_\gamma \propto \rho_\phi$,
and so the condition for warm inflation to end is never met. In this
case warm inflation continues until the slow-roll conditions are 
no longer met and the inflaton field decays away, 
which then ends inflation.
For $n > 4$ and exponential potentials, the radiation density will always 
fall faster than the energy-density of the inflaton field (see Appendix
B). In this case warm inflation will occur, in the sense
that a sizable radiation component can persist for
a considerable duration during inflation, well beyond the requirements
for solving the problems of standard cosmology.  However
to end such a warm inflation regime, this behavior of the potential
must be local, and at some point the vacuum energy must be released 
at a sufficiently fast rate to allow the radiation to dominate.

\section{Specific Cases}

In order to understand how warm inflation differs from the standard
supercooled inflationary 
scenario, it is useful to consider a few specific examples.
Here we consider models of warm inflation with polynomial 
and exponential potentials.

\subsection{Polynomial inflation}

We consider the simplest inflationary model
\be
		V(\phi)= m^2 \phi^2 /2
\label{n=2}
\ee
for a massive scalar field with mass $m$.
With this potential the slow-roll parameters are
\be
	\e = \eta = \frac{m_p^2}{4 \pi \phi^2}.
\ee
The Hubble parameter is given by
\be
	H(\phi)=\sqrt{\frac{4\pi}{3}}\left( \frac{m}{m_p}\right) \phi
\ee
while the dissipation parameter scales like
\be
	r = \frac{1}{\sqrt{12\pi}}\left( \frac{\Gamma}{m}\right)
		\frac{m_p}{\phi}.
\label{rn=2}
\ee
From equation (\ref{rn=2}) we see that 
the ratio $\Gamma/m$ controls the evolution of the system.

Solving equation (\ref{phidot2}) we find that the inflaton field evolves as
\ba
	\phi(t) &=& \phi_0 e^{-m^2 t/\Gamma} \nn	
		&\approx& \phi_0 \left(1- \frac{m^2}{\Gamma} t \right).
\label{phisol}
\ea
Compared with standard inflation, where \cite{liddle}
\be
	\phi(t) \approx \phi_0 - \left( \frac{m m_p}{\sqrt{12 \pi}} \right) t,
\ee
we see that during warm inflation the inflaton field decays due to 
dissipation into the radiation field.

The energy densities of the inflationary potential 
is given by equation (\ref{inflden}) 
and (\ref{n=2}) while the radiation field is given by 
\be
	\rho_\gamma=\sqrt{\frac{3}{\pi}}
	\left( \frac{m^3 m_p}{8 \Gamma}\right) \phi
	= \sqrt{\frac{6}{\pi}}\left( \frac{m^2 m_p}{8 \Gamma}\right) 
	\rho_\phi^{1/2}.
\label{anngam}
\ee
The number of e-folds from the end of warm inflation  is given by
\ba
	N(\phi) &=&  -\ln k(\phi) \nn 
	&=& - \frac{8 \pi}{m^2_p} \int_\phi^{\phi_{\rm end}} 
	d \phi' \frac{rV}{V'} \nn
	&=& \sqrt{\frac{4\pi}{3}}\left(\frac{\Gamma}{ m}\right) \frac{1}{m_p}
	(\phi - \phi_{\rm end}).
\ea

Figure 1 shows the evolution of the energy densities of the radiation
and inflationary fields as a function of
number of e-folds. The upper line is the inflation energy density and
the lower line the radiation energy density. The dashed lines are 
the results of our analytic model, from equations (\ref{n=2}), (\ref{anngam})
and (\ref{efoldn=2}), while the solid lines are the 
results of our numerical model (Section \ref{num}) of the evolution 
of the field through the warm inflationary era.

\begin{figure}[t]
\vspace{-3.cm}
 \centerline{\epsfig{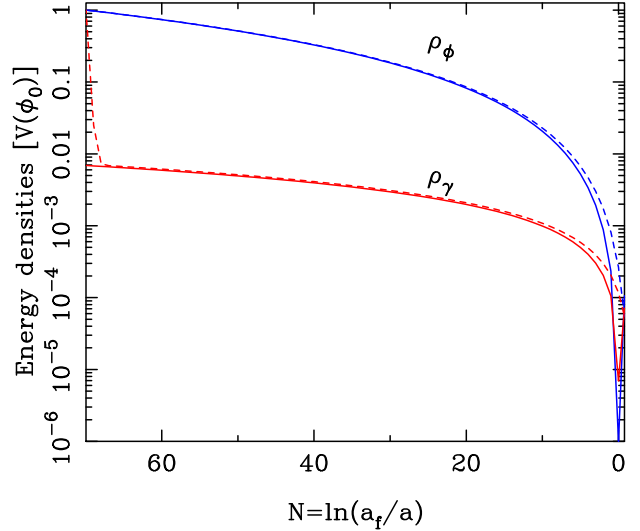}}
\vspace{-3.cm}
 \caption{The evolution of energy density of the inflationary and radiation 
fields as a function of number of e-folds from the end of warm inflation, 
$N=\ln (a_f/a)$, where $a_f$ is the final expansion factor, 
for the inflationary potential $V=m^2\phi^2/2$. The energy-density of the 
inflationary field (upper curves) and the radiation field (lower curves) 
are both normalised to unity in units of the magnitude of the potential 
at the start of inflation, $V(\phi_0)$. The solid lines are the results 
of the analytic expressions equations (\ref{anngam}), while the dashed 
lines are the results of the numerical model discussed in Section \ref{num}. }
\end{figure}

From the condition that $\e=2r$ at the end of warm inflation we find that
the magnitude of the inflationary field at this time is
\be
	\phi_{\rm end} = \frac{1}{2} \sqrt{\frac{3}{4 \pi}} m_p
	\left(\frac{m}{\Gamma} \right)
\ee
where the initial field must have $|\phi_0| > \phi_{\rm end}$. In standard
polynomial $(n=2)$ inflation this condition is 
$|\phi|_{\rm end}>m_p/\sqrt{4 \pi}$,
so we find that for warm inflation this is weakened by the factor 
$\approx m/\Gamma$.
If $\Gamma > m$ then warm inflation can begin well below the Planck mass. 

The total number of e-folds from the end of warm inflation is
\be
	N_{\rm tot} = \sqrt{\frac{4\pi}{3}} \left( \frac{\Gamma}{m}\right)
		\left( \frac{\phi_0}{m_p}\right) -\frac{1}{2}.
\label{efoldn=2}
\ee
As we require $N\approx 70$ e-folds to comfortably solve the problems 
of standard cosmology this implies
\be
	\phi_0 \approx 35 \left( \frac{m}{\Gamma} \right) m_p.
\label{phi0}
\ee

\begin{figure}[t]
\vspace{-3.cm}
 \centerline{\epsfig{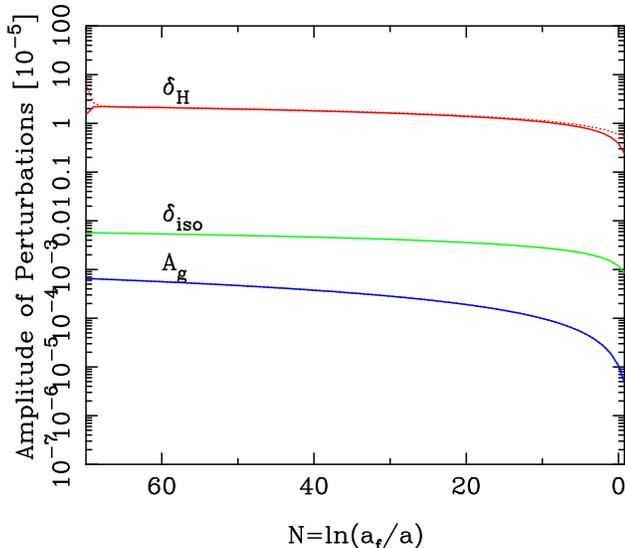}}
\vspace{-3.5cm}
 \caption{The perturbation spectra of adiabatic scalar perturbations 
(top line),
isocurvature scalar perturbations (middle line) and 
gravitational waves (bottom line) as a function of e-folds from the
end of warm inflation, $N$, for the potential 
$V=m^2 \phi^2/2$. The model is normalised to have 
$\delta_H = 2 \times 10^{-5}$ at the $50^{th}$ e-fold.
Solid lines are the results of our analytical model, while
the dotted lines are from the numerical model.}
\end{figure}

The requirement of strong dissipation also leads to a constraint on 
$\Gamma$. From equation (\ref{rn=2}) with equation (\ref{phi0}) we
find that at the beginning of inflation 
\be
	r(\phi_0) \approx \frac{1}{215} \left( \frac{\Gamma}{m} \right)^2 \gg 1
\ee
resulting in the requirement that $\Gamma \gg 15 m$ for warm inflation
to commence in polynomial ($n=2$) inflation.

Calculating the spectrum of adiabatic perturbations yields 
\be
	\delta_H^2 = \frac{2}{75} \left( \frac{2}{3 \alpha \pi^5}\right)^{1/4}
	 	\left(\frac{\Gamma}{m_p}\right)^{3/2} 
		\left(\ln \frac{k}{k_{\rm end}} \right)^{3/4},
\ee
where $\ln k_{\rm end}=\sqrt{4 \pi/3}\Gamma \phi_{\rm end}/(m m_p)
\approx 1/2$ is the 
fiducial wavenumber at the end of warm inflation. Note that the 
amplitude of adiabatic perturbations
is independent of the magnitude of the inflationary potential.
In Appendix B we show that the amplitude of adiabatic perturbations
in warm inflation is independent of the magnitude of the inflationary 
potential for all polynomial potentials.

Figure 2 shows the amplitude of adiabatic, isocurvature and tensor
perturbations for the $n=2$ polynomial. In Appendix B we derive
results for the more general polynomial and exponential cases.
We find that the ratio of the amplitude of perturbations agrees with the
inequality expressed by equation (\ref{ratioineq}). For the $n=2$ case the 
isocurvature and tensor perturbations are strongly suppressed
below the adiabatic perturbations. This difference can be reduced 
by increasing the magnitude of the inflationary potential.

\begin{figure}
\vspace{-3.cm}
 \centerline{\epsfig{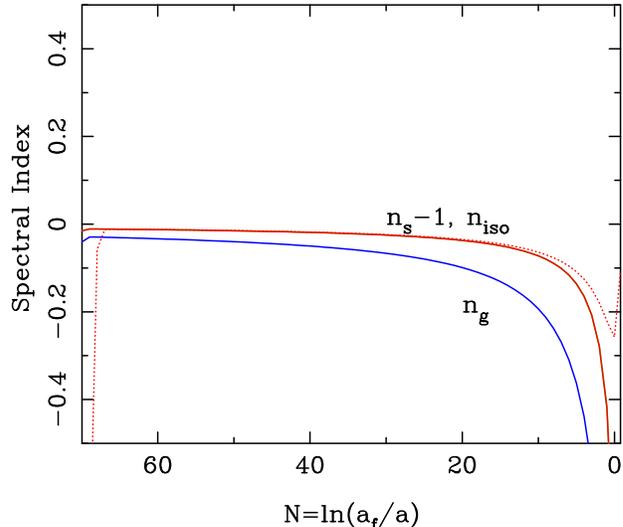}}
\vspace{-3.5cm}
 \caption{The spectral indices of the adiabatic scalar perturbations, $n_s-1$, 
isocurvature scalar perturbations, $n_{\rm iso}$, and 
gravitational waves, $n_g$, as a function of e-folds from the end
of warm inflation, $N$, for the potential 
$V=m^2 \phi^2/2$. The solid lines are for our analytic  model, while dotted
lines are from the numerical model. The spectral index of the isocurvature
perturbations lies under that of the adiabatic curve.}
\end{figure}

The spectral indices are related by
\be
	n_s-1 = n_{\rm iso} = \frac{3}{8} n_g = -\frac{3\e}{4r},
\ee
and are all red.
In Appendix B we derive more results for the general 
case of $\phi^n$ potentials. There we show that for polynomial potentials
the spectral index is always red unless $n>4$, or for the case of
an exponential potential (see Appendix B), in which case the
spectrum is blue.  The latter open up the interesting
possibility of primordial black 
hole production in the early universe \cite{blue}.
A scale invariant spectrum, $n_s=1$, is found only for the $n=4$ model.

The dissipation ratio, $r$, and the slow-roll parameter $\e$ are 
related by
\be
	r = \frac{1}{\sqrt{3}} \left( \frac{\Gamma}{m} \right) \e^{1/2}
\ee
so that the ratio of isocurvature to adiabatic
perturbations, given by $\e/r^2$, is
$3(m/\Gamma)^2 \ll 1$. 

%Hence in the $n=2$ polynomial warm inflation model
%isocurvature modes are negligible.

Figure 3 shows the spectral indices of the adiabatic, isocurvature and
tensor spectra as a function of wavenumber. In all cases the 
spectral indices starts as a flat spectrum, when $\e \ll r$. 
In the case of 
$n=4$ spectral indices are exactly flat while for the exponential potential 
the spectral indices are almost flat.

\section{Numerical Models}
\label{num}

While our analytic results simplify the whole analysis of warm inflationary
models, it is important to verify the range of validity of this analysis
and find 
where the assumptions used in their derivation are accurate. 
To test this we have numerically
evolved the inflationary and radiation fields through a warm
inflationary epoch, and calculated the adiabatic and tensor perturbation
spectra. We numerically solve equations (\ref{evolvphi}) and (\ref{evolvgam})
for the evolution of the field and equation (\ref{fried_rad}) for the 
evolution of the background universe using the methods of \cite{ab2}.
 
The results for the evolution of
the energy-density of the fields for the case of a polynomial $n=2$
potential are shown in Figure 1. The magnitude of the inflationary 
and radiation energy-density fields are both normalised to unity in units 
of the potential at the start of inflation. We set $\rho_\gamma=\rho_\phi$
initially, as it is reasonable to assume that inflation begins from a
radiation-dominated epoch. We tested a number of other initial conditions
including $\rho_\gamma=0$ and found little variation in our final 
results, as the models soon settled down to the same evolution after
the first few e-folds.

To parameterise the models we used the
more general expression for polynomial inflation from Appendix B,
and choose parameters so that the amplitude of adiabatic perturbations
$\delta_H = 2 \times 10^{-5}$ at the $50^{th}$ e-fold, and allowed 
a total of 70 e-folds of warm inflation to elapse. 

Figure 1 verifies that not only the shapes of the curves are well 
approximated, but also the relative magnitudes of the radiation and 
inflationary 
energy-densities are accurately reproduced. The main assumptions 
underlying our analytic results are that of
slow-roll inflation and quasi-stable radiation production 
($\dot{\rho}_\gamma \ll \Gamma \dot{\phi}^2$).
The latter assumption breaks down at the start of warm inflation
as the radiation field initially is redshifted away. But radiation
production by dissipation of the inflaton field soon catches up with 
the redshifting and stability is reached after a few e-folds.

We found accurate agreement between our analytic models and the 
full numerical models for polynomial potentials in the range 
$0.5<n<4$. In most cases the agreement was $<1\%$ over most of the 
range, except near the beginning of inflation.
In the $n=4$ case the energy densities tracked each other
as expected from our discussion in Section IV. We also tested our results
for the exponential potential discussed in Appendix B, and again 
found similar accuracy, although here again the radiation field 
drops faster than the energy density in the inflationary field and
warm inflation has to end by the natural decay of the inflaton,
or by some other means.

We have also computed the amplitude and scale dependence of the perturbation
spectra for adiabatic perturbations directly from equations 
(\ref{adpert}) and (\ref{dphi}). These are plotted for the case
of $n=2$ in Figure 2. Again close agreement was found,
with deviations less than $1\%$ over the range of wavenumbers. Again
the poorest agreement was over the few e-folds at the beginning and 
end of warm inflation.

Finally we tested the predicted spectral indices for adiabatic 
perturbations (Fig. 3). Again we found good agreement between  
the analytic and numerical models, except for the first and last
few e-folds. The sudden change in $n_s-1$ at the beginning of
warm inflation ($N \approx 70$) is real and due to the sharp 
transition to warm inflation when the change in the radiation density,
$\dot{\rho}_\gamma$, is important.

\section{Observational Constraints}

In standard inflation, there are four observable constraints ($\delta_H$, 
$A_g$, $n_s$, $n_g$) for the three parameters of slow-roll inflation,
$V_0$, $\e$, $\eta$, where $V_0$ is the magnitude of the inflationary 
potential. With four constraints there is a redundancy in the observations
allowing a consistency relation \cite{liddlelyth}. This is usually 
expressed as a relationship
between the tensor-to-scalar ratio and the slope of the tensor spectrum,
\be
	R_g = - \frac{n_g}{2}.
\ee
This appears to be generally true for slow-roll inflation, although
recent work on pre-heating at the end of supercooled inflation 
may effect this result \cite{bassett,brand,ep,llmw}.

In warm inflation the dissipation factor, $r$, becomes an active parameter
(since it is present at the end of supercooled inflation, but assumed inactive)
giving four constraints for four parameters. Hence we do not expect the
consistency relation of standard inflation to hold in warm inflation.
This relation can only arise by coincidence. If the consistency 
relation is not observed it may indicate a warm inflationary phase.
The dissipation factor can be estimated from the observables by the relation
\be
	r^2 \approx 
	-n_g \left(\frac{25 \alpha \pi^2}{32} R_g^4 A_g^2\right)^{1/3}.
\ee

To discriminate between warm and standard inflation requires a measurement of
all four observables. The MAP and Planck satellite missions will probe
the CMB and provide strong constraints on the amplitude and shape of 
the scalar perturbation spectrum. Both MAP and Planck have polarisation 
detectors which are sensitive to tensor modes, but Planck has a much
broader range of frequencies which will aid the subtraction of foreground
contamination of the CMB signal.
In practice the tensor modes may be so small as to be unobservable,
hence it may not be easy to distinguish between the two models. 

Current observations, from the COBE satellite, and subsequent 
airborne detectors (see \cite{tegzald} for a recent analysis), can be used 
to place constraints on the amplitude of 
scalar perturbations and the slope of their spectra.
The low COBE multipoles indicate that $\delta^2_H \approx 10^{-10}$, 
giving the constraint 
\ba
	V^{1/4} &\approx& 7.6 \times 10^{-4}
	 (\alpha^{1/3} \e r^{-3})^{1/4} m_p, \nn
	&\approx& 7.6 \times 10^{15}(\alpha^{1/3} \e r^{-3})^{1/4} {\rm GeV}.
\ea
If $r^3 \approx \alpha^{1/3} \e$, these energy scales are pretty 
much in keeping 
with those of Grand Unified Theories. However if $r\gg 1$ as required
for warm inflation we open up the interesting possibility that the
energy scale required can be much lower.

\section{Conclusions}
 In this paper we have investigated the phenomenology of the warm 
inflationary scenario. In the slow-roll, warm inflation epoch we have
found the general relationship between the energy-density in the
inflaton and radiation fields. This has led us to the general criteria
for warm inflation, $\e<r$, $|\eta|<3r^2$ and $r>1$. The usual 
inflation condition, $\e<1$, is generalised as dissipation can 
help sustain accelerated expansion of the universe for a wider range
of inflationary potentials. 

Generalising the analysis of inflation in 
the slow-roll regime, we have derived expressions for the perturbation
spectra of adiabatic and isocurvature fluctuations generated by thermal 
fluctuations during the warm inflationary epoch. The adiabatic spectrum 
produced by thermal fluctuations dominates that 
produced by quantum fluctuation.
Its amplitude and spectral slope are both different from
quantum generated perturbations and can lead to red or blue spectra.
Isocurvature perturbations also are generated due to internal 
fluctuation between radiation fields. These are generally suppressed
compared to the adiabatic spectrum.

Gravitational waves are still generated via quantum excitation, and
have the same amplitude but different slope from standard inflation. The 
tensor-to-scalar ratio is suppressed by the relatively high amplitude of
the adiabatic spectrum. 

No consistency relation exists in warm inflation as there are now four
parameters (the standard three of inflation and an active dissipation factor)
and in general only four observables. This is probably the most 
important observational test of warm inflation, but may be difficult,
even for Planck, if the tensor modes are strongly suppressed. In this 
case measurement of the isocurvature modes relative to adiabatic would
provide an alternative test. However,
verification of the consistency relation would rule out substantial 
interactions during inflation.

There are two important differences between standard and warm inflation. 
First, the 
slope of the inflationary potential does not need to be small, as dissipation
will allow slow-roll even for steep potentials. Second, the energy scale 
of warm inflation may be much lower than required for standard inflation,
due to the higher amplitude of thermal adiabatic perturbations and the 
constraints of Sachs-Wolfe effects observed in the CMB.

To investigate the warm inflationary model we need to study further
 the evolution of scalar perturbations during the transition 
phase from warm inflation to radiation domination when the energy
densities of the radiation field and inflationary field are comparable.
This will occur more smoothly than in standard
inflation and should lead to observational differences between the
models. In addition the model we have presented here for isocurvature
perturbations is crude. A more complete understanding of the generic
prediction of isocurvature perturbations for warm inflation is needed
before detailed predictions can be made.

The model we have presented here is an extreme one, in that we have 
assumed that it is dissipation--dominated. A whole range of models
lie between that of warm inflation and the standard supercooled
inflation, and it would be interesting to explore their properties
in the general framework we have outlined here. The development in \cite{maia}
also would be beneficial for this.

Finally these results 
should be propagated to the post-recombination era using 
general-purpose transfer codes such as CMBfast \cite{seljakzald}.
However to make definite, high-quality predictions for the CMB, such 
codes must be generalised to remove any of the relations of standard 
supercooled inflation, such as the consistency relation. This will
allow us to predict with some confidence whether the MAP and Planck
missions will be able to probe the high redshift universe 
with sufficient accuracy to distinguish between warm and supercooled 
inflation.

Acknowledgements: ANT and AB acknowledge the support of PPARC, and thank
David Wands for useful discussion about the evolution of non-adiabatic
super-horizon perturbations.

\section*{Appendix A}

The origin of the ``freeze-out'' wavenumber $k_F$ in equation (\ref{dphi})
is reviewed here following the original and more detailed explanation
in \cite{abden}.  
To account for fluctuations,
the inflaton now is written as 
$\phi(\x,t) = \phi_0(t) + \delta \phi(\x, t)$.
Assuming the universe remains near thermal equilibrium
during warm inflation, the fluctuations of the inflaton field
obey a fluctuation-dissipation relation.  
The evolution equation for the fluctuations,
$\delta \phi (\x, t)$, is obtained from the linearized deviation
of the zero mode equation of motion equation (\ref{evolvphi}), with
inclusion of the spatial Laplacian term and the addition
of a white--noise random force term
\begin{equation}
\Gamma 
\frac{d \delta \phi(\k,t)}{dt} = 
-[{\bf k}^2+V''({\phi_0})] \delta \phi(\k,t)
+\xi(\k,t),
\label{eomdelp}
\end{equation}
where $V''(\phi_0) = d^2V(\phi_0)/d \phi_0^2$ and
we have Fourier transformed to momentum space. 
The white--noise represents the action of the thermal heat bath on
the inflaton field, with the spectrum of the this noise term
governed by the fluctuation-dissipation theorem, 
\begin{equation}
\lgl \xi (\k,t)\rgl =0
\end{equation}
and
\begin{equation}
\lgl \xi(\k,t) \xi(-\k^{\prime},t^{\prime}) \rgl_\xi 
\stackrel{T\rightarrow \infty}{=}
2 \Gamma T
(2\pi)^3 \delta^{(3)}(\k-\k^{\prime}) \delta(t-t^{\prime}).
\end{equation}
The solution of equation (\ref{eomdelp}) is
\begin{eqnarray}
\delta \phi(\k,t) & \approx & \frac{1}{\Gamma}
e^{-(t-t_0)/\tau(\phi)}
\int_{t_0}^{t}
e^{(t'-t_0)/\tau(\phi_0)}
\xi(\k,t') dt' \nonumber\\
& + &
\delta \phi(\k,t_0)
e^{-(t-t_0)/\tau(\phi_0)},
\label{delpsol}
\end{eqnarray}
where $\tau(\phi)=\Gamma/(k^2 + V''(\phi))$.
The first term on the right hand side acts to thermalize
$\delta \phi$, whereas the latter is the memory term for
the initial value of $\delta \phi$, which  over time becomes
negligible.
 
In the cosmological case, equation (\ref{eomdelp}) must be interpreted
for physical wavenumbers, since the thermal effects as expressed
in this equation act in accordance with the physical wavenumbers
of the system.
Thus for a given fluctuation mode, $\delta \phi(\k_c)$, at comoving
wavenumber $\k_c$,
equation (\ref{eomdelp}) expresses its dynamics within
the time interval $1/H$ when the physical
mode is 
$k_{phys} \equiv k_{c} {\rm e}^{-Ht} \approx k$.
To be consistent with the thermalization conditions,
during the time interval $\sim 1/H$ the mode
$\phi(\k_c)$ must thermalize at physical scale $k$.
In terms of the solution equation (\ref{delpsol}), this requires the memory
term to be negligible within a Hubble time, thus
\begin{equation}
\frac{k^2 + V''(\phi_0)}{H \Gamma} > 1.
\end{equation}
The freeze-out wavenumber $k_F$ is at the point where this condition
first holds, which for $V''(\phi_0) < \Gamma H$ is 
\begin{equation}
k_F = \sqrt{\Gamma H}.
\label{kf}
\end{equation}
If $V''(\phi_0) > H \Gamma$ then the freeze-out wavenumber
is the same as for supercooled inflation, $k_F=H$. However, in general
this regime never occurs during warm inflation and equation (\ref{kf})
is the governing condition.
Note that $H$ varies during warm inflation, thus based
on equation (\ref{kf}), so too will $k_F$. However, the variation of
$H$ is typically very small, less than a factor 10, during the entire
warm inflation period.  Thus, up to order one factors,
one can treat $k_F$ as a constant
with $H$ evaluated at some appropriate time during warm inflation,
such as the beginning. In our analytic approach we have included the 
$\phi$-dependence of $k_F$.

In our treatment here we have taken the dissipative
coefficient $\Gamma$ to be independent of $k$, whereas
in \cite{abden} it was mentioned that in general $\Gamma$
will depend on $k$ and generally will decrease
with increasing wavenumber.  In this case, the condition
equation (\ref{kf}) still is valid except it must now be solved
for $k_F$ due to the $k$-dependence in $\Gamma$.

\section*{Appendix B}
%\subsection*{Polynomial warm inflation}
\subsection*{Polynomial potentials}

In this appendix we explore the properties of the warm inflationary scenario
for the family of models with the functional form
\be
	V(\phi) = \lambda m^4 \left( \frac{\phi}{m} \right)^n
\label{expopot}
\ee
where $\lambda$ is a dimensionless parameter, which 
allows more freedom in 
the normalisation of the potential.
 
The slow-roll parameters are given by
\be	
	\e = \frac{n^2 m_p^2}{16 \pi \phi^2}, \hspace{0.5in}	
	\eta = \frac{n(n-1)m_p^2}{8 \pi \phi^2}
\ee
with the relation
\be
	\e = \frac{n}{2(n-1)} \eta.
\ee
The expansion of the universe is governed by the Hubble parameter
\be
	H(\phi) = \sqrt{\frac{8 \pi\lambda}{3}} \frac{m^2}{m_p} 
	\left( \frac{\phi}{m}\right)^{n/2},
\ee
where the dimensionless dissipation parameter is
\be
	r = \frac{1}{\sqrt{24 \pi \lambda}} \left( \frac{\Gamma}{m}\right)
	\left( \frac{m_p}{m}\right)\left(\frac{\phi}{m} \right)^{-n/2}.
\ee
The energy density of the radiation field is given by
\be
	\rho_\gamma = \sqrt{\frac{3\lambda^3}{2\pi}} \frac{n^2}{8} 
	\frac{m_p}{\Gamma}
	 m^4 \left(\frac{m}{\phi} \right)^{(4-3n)/2}, 
\ee
and related to  the energy-density of the inflation field by
\be
	\rho_\gamma = \sqrt{\frac{3}{2\pi}} \frac{n^2}{8}  
	\frac{m_p}{\Gamma} m_p^4
	\left[ \lambda \left(\frac{m}{m_p} \right)^{4-n} \right]^{2/n} 
	 \left(\frac{2 \rho_\phi}{m_p^4}\right)^{\frac{(3n-4)}{2n}} .
\label{radenergy}
\ee
This last relation tells us that for an $n=4$ potential the energy densities 
are proportional, $\rho_\gamma \propto \rho_\phi$, and that warm inflation
cannot end by the condition given by equation (\ref{endwi}). When the
potential is steeper than this, $n>4$, or for the exponential potential, 
the radiation field drops more rapidly than the inflation field.
Although warm inflation occurs, the effects of the radiation field
will diminish and inflation will begin to evolve in the same 
way as for standard inflation. Hence for polynomial inflationary potentials we
require $n<4$ for strong warm inflation to take hold.

The number of e-folds, $N(\phi)$, the universe expands during inflation is
\be
	N(\phi) = \sqrt{\frac{2\pi}{3\lambda}} \frac{8\Gamma}{n(4-n)m_p}
	\left[\left( \frac{\phi_0}{m}\right)^{\frac{(4-n)}{2}}-
	\left( \frac{\phi}{m}\right)^{\frac{(4-n)}{2}} \right]
\label{efolds}
\ee
where we count the number of e-folds from the start of warm inflation
(note in Section V we counted e-folds from the end of warm inflation).
 The number of e-folds can be related to $\e$ and $r$ by
\be
	N(\phi)= \left( \frac{2n}{4-n} \right) \left(\frac{r_0}{\e_0} 
	-\frac{r}{\e} \right).
\ee
If $n<4$ then $N$ is dominated by the ratio of $\e/r$ at the start 
of the inflationary phase, and to a good approximation the total 
number of e-folds is
\be
	N_{\rm tot} = \left( \frac{2n}{4-n} \right)\frac{r_0}{\e_0}.
\ee

The adiabatic spectrum of density perturbations is
\be
	\delta_H^2 = \frac{8}{75} 
	\left( \frac{2^7}{3^5\pi^7\lambda^3}\right)^{1/8}
	\left( \frac{\Gamma^9}{\alpha n^6 m_p^9} \right)^{1/4}
	\left( \frac{\phi}{m} \right)^{3(4-n)/8}.
\ee
Substituting in equation (\ref{efolds}) we find that the amplitude
of adiabatic perturbations in warm inflation is independent of the
magnitude of the inflationary potential for polynomial potentials,
when expressed as a function of e-folds, or scale.

The spectral index of adiabatic fluctuations is related to the slow-roll
parameter, $\e$, and the dissipation parameter, $r$, by
\be
	n_s -1 = -\frac{(4-n)}{n} \frac{3\e}{4r} = 
	\frac{(4-n)}{n} \frac{3}{8} n_g 
	=\frac{3(4-n)}{(5n-4)} n_{\rm iso}.
\ee
where the last two expressions relate the adiabatic spectral index to the 
isocurvature and tensor spectral index. From this expression we see that the 
warm inflationary adiabatic spectrum is only exactly scale invariant
($n_s=1$) for potentials with $n=4$, and is red unless $n>4$.

The isocurvature spectrum is given by 
\be
	\delta_{\rm iso}^2 = \frac{1}{50} 
	\left( \frac{\lambda^5}{6^3\pi^{15}}\right)^{1/8}
	\left( \frac{m}{m_p} \right)^2
	\left[ \frac{n^2}{\alpha} \frac{\Gamma}{m_p}
	\left( \frac{\phi}{m}\right)^{(5n-4)/2}\right]^{1/4}.
\ee

The ratio of the amplitude of isocurvature to adiabatic perturbations 
is given by
\be
	R_I = \frac{\e}{16 \pi r^2} = 
	\frac{3\lambda n^2}{32 \pi} 
	\left( \frac{m}{\Gamma} \right)^2
	\left(\frac{\phi}{m} \right)^{n-2}.
\ee
In the special case of $n=2$ this is a constant, as found in Section V.

\subsection*{Exponential potential}
Consider potentials of the form 
\be
	V(\phi) = V_0 \exp  \frac{\phi}{\kappa},
\ee
where $\kappa$ is a free parameter with units of mass.

The slow-roll parameters are 
\be
	\eta = 2\varepsilon=\frac{ m_p^2}{8 \pi \kappa^2} .
\ee

The energy densities of the radiation and inflationary fields are 
related by
\be
	\rho_\gamma = \sqrt{\frac{3}{2 \pi}} 
	\left(\frac{ m_p}{8 \Gamma \kappa^2}\right)
	\rho_\phi^{3/2}.
\ee
Hence the energy density of the radiation field falls more rapidly than
that of the inflationary field, as predicted. In the exponential model, 
warm inflation will end in the same way as discussed earlier 
for $n>4$ polynomial potentials.

The number of e-folds from the start of warm inflation is
\ba
	N(\phi) &=& \sqrt{\frac{2\pi}{3}} 
	\left(\frac{4 \Gamma \kappa^2}{ m_p}\right)
		(V^{-1/2}(\phi_0)-V^{-1/2}(\phi)) \nn
		&=& \left( \left( \frac{r}{\e}\right)_0 - 
		\left( \frac{r}{\e}\right)_\phi \right).
\ea

The amplitude of adiabatic scalar perturbations is
\be
	\delta_H^2 = \frac{8}{75} \left(\frac{2}{\pi}\right)^{7/8}
	\frac{1}{3^{5/8}\alpha^{1/4}} \left(\frac{\kappa}{m_p} \right)^{3/2}
	 \left(\frac{\Gamma}{m_p}\right)^{9/4} 
	\left(\frac{V}{m^4_p}\right)^{-3/8}
\ee
while that of isocurvature perturbations is given by
\be
	\delta^2_{\rm iso} = \frac{1}{50} (6^3 \pi^{15})^{1/8}
	\left(\frac{\Gamma}{m_p} \right) 
	\left(\frac{ m_p^2}{a\kappa^2}\right)^{1/4}
	\left( \frac{V}{m_p^4} \right)^{5/8}
\ee
with ratio
\be
	R_{\rm iso} = \frac{9 \pi^{11/4}}{16 \sqrt{2}}
	\left( \frac{\Gamma}{m_p}\right)^{-2} 
	\left(\frac{m_p}{\kappa}\right)^{2}
	\left( \frac{V}{m_p}\right).
\ee

The adiabatic and isocurvature spectral indices are 
\be
	n_s-1=\frac{3\varepsilon}{4r}, \hspace{0.5in}
	n_{\rm iso} = -\frac{5 \e}{4r}.
\ee
For warm inflation with an exponential potential,
the adiabatic spectrum is always blue
in contrast with  standard inflation while the isocurvature is always red.
The spectral indices are related by
\be
	n_s-1 = - \frac{3}{5} n_{\rm iso}=-\frac{3}{8}n_g.
\ee

\end{document}